\documentclass[english]{article}
\usepackage[utf8]{inputenc}
\usepackage{amsmath}
\usepackage{amssymb}

\usepackage{mathrsfs}
\usepackage{dsfont}
\usepackage{csquotes}
\usepackage{stackengine}
\usepackage{amsmath,amssymb}
\usepackage{mathtools}
\usepackage{hyperref}
\usepackage[normalem]{ulem} %for \sout{}
\usepackage{color}

\usepackage[
    backend=biber,
    natbib=true,
    style=ieee,
    url=false,
    sorting=none,
    maxbibnames=6,
    minbibnames=6,
    doi=false
]{biblatex}
\bibliography{references.bib}
\allowdisplaybreaks

\makeatletter

\def\beq{\begin{equation}}
\def\eeq{\end{equation}}
\def\bsp#1\esp{\begin{split}#1\end{split}}
\def\bal#1\eal{\begin{align}#1\end{align}}

\newcommand{\bzh}{\hat{\textbf{z}}}

\DeclareMathOperator{\Tr}{Tr}
\newcommand{\rd}{\textrm{d}}
\newcommand{\cM}{\mathcal{M}}
\newcommand{\bq}{\textbf{q}} 
\newcommand{\bs}{\textbf{s}}

\newcommand{\bb}{\textbf{b}} 
\newcommand{\ba}{\textbf{a}} 
\newcommand{\bd}{\textbf{d}} 
\newcommand{\bB}{\textbf{B}} 
\newcommand{\bC}{\textbf{C}} 
\newcommand{\bQ}{\textbf{Q}} 
\newcommand{\bS}{\textbf{S}} 
\newcommand{\blambda}{\lambda}

\begin{document}
%%%
%%% Title page
%%%

\begin{titlepage}

% Title
\noindent
BONN-TH-2025-23
\vspace{1cm}
\begin{center}
{\LARGE \bf 
A critical appraisal of tests of locality and of entanglement versus non-entanglement at colliders
}
\vskip 1cm

\large
Philip Bechtle$^a$, Cedric Breuning$^a$, Herbi K. Dreiner$^{a,b}$, Claude~Duhr$^{a,b}$\\
\vspace{0,5cm}
\normalsize
{\it $^{\, a}$ Physikalisches Institut, Universität Bonn, D-53115 Bonn, Germany\\
$^b$Bethe Center for Theoretical Physics, Universit\"at Bonn, D-53115 Bonn, Germany}\\
\vspace{0,5cm}
Emails: \texttt{bechtle@physik.uni-bonn.de}, \texttt{ccbreuning@uni-bonn.de}, \texttt{dreiner@uni-bonn.de}, \texttt{cduhr@uni-bonn.de}
\vspace{1.4cm}

{\large \bf Abstract}
\vspace{-0.2cm}
\end{center}
It has been argued more than  30 years ago that it is not possible to test locality at colliders, due to the inability to directly measure non-commutating observables such as spin components in current collider experiments. Recently, there has been a lot of phenomenological and experimental activity around testing locality via Bell-type experiments or entanglement versus non-entanglement in a collider environment. These results seem to evade the earlier no-go theorem by indirectly measuring spin correlations via their relation to angular correlations between momenta. We perform a careful study of the feasibility of such an approach. We scrutinize the relationship between spin and angular correlations in both quantum mechanics and local hidden variable theories. Our conclusion is that it is currently not possible to perform a logically coherent set of experimental measurements at colliders that would allow one to test locality or entanglement versus non-entanglement. This reaffirms the earlier no-go theorem. We stress that the no-go theorem does not apply to measurements of observables inspired from entanglement and Quantum Information Theory to test the Standard Model of particle physics.
\end{titlepage}
\clearpage

%%%
%%% TOC here
%%%

\tableofcontents

%%%%%%%%%%%%%%%%%%%%%%%%%%%%%%%%%%%%%%%%%%%%%%%%%%%%%%%%%%%%
%%%% Sections
%%%%%%%%%%%%%%%%%%%%%%%%%%%%%%%%%%%%%%%%%%%%%%%%%%%%%%%%%%%%

% !TEX root = main.tex

\section{Introduction}
\label{sec:intro}

One of the most striking features that sets quantum mechanics (QM) apart from classical theories is entanglement, which leads to correlations between outcomes of causally-disconnected experiments that cannot be explained by classical local theories. For this reason, entanglement and its consequences have been the subject of interest since the early days of QM, cf. ref.~\cite{Einstein:1935rr}. 
It was realized~{\cite{Bell:1964kc}} that QM leads to results that are inconsistent with certain \emph{local hidden variable theories} (LHVTs),\footnote{See appendix~\ref{app:LHVTs} for a brief review {of LHVTs}.}. {In these LHVTs} a system is described by a set of \emph{hidden} variables ${\blambda}=(\lambda_1,\ldots,\lambda_n)\in \Lambda$ which determine the outcome of the experiments, but which remain themselves inaccessible to direct experimental measurements. Instead, the hidden variables are distributed according to a certain probability density function (PDF) $P({\blambda})$, and the expectation value of an observable {$A$} measured in an experiment takes the form
\beq\label{eq:A_LHVT}
\langle A(\ba)\rangle = \int_{\Lambda}\rd^n\blambda\,A(\ba,\blambda)\,P(\blambda)\,.
\eeq
Here $\ba=(a_1,\ldots,a_m)$ denote the settings of the experiment, which are assumed be 
to independent of the hidden variables $\lambda$. {In this context} \emph{locality} 
refers to the fact that the outcomes of a measurement of another observable $B(\bb)$ that is causally disconnected from $A(\ba)$ cannot be influenced by the outcomes of a measurement of $A(\ba)$. In particular, this implies that the settings $\bb$ must be independent of $\ba$, and the correlation function of the product takes the form
\beq\label{eq:AB_LHVT}
\langle A(\ba)B(\bb)\rangle = \int_{\Lambda}\rd^n\blambda\,A(\ba,\blambda)\,B(\bb,\blambda)\,P(\blambda)\,.
\eeq

The distinctive features of LHVTs compared to QM were quantified in the famous paper by Bell~\cite{Bell:1964kc}, who showed that, in the context where the observables $A$ and $B$ take the discrete values $\pm1$, the correlation function in eq.~\eqref{eq:AB_LHVT} necessarily satisfies  certain inequalities which may be violated in QM. In particular, Clauser, Horne, Shimony and Holt (CHSH) showed that, as a consequence of Bell's result, in any LHVT the correlation function in eq.~\eqref{eq:AB_LHVT} satisfies the inequality~\cite{CHSH}
\beq\label{eq:CHSH}
\big|\langle A(\ba)B(\bb)\rangle -\langle A(\ba')B(\bb)\rangle +\langle A(\ba)B(\bb')\rangle +\langle A(\ba')B(\bb')\rangle\big|\le 2\,.
\eeq
In QM the CHSH inequality in eq.~\eqref{eq:CHSH} may be violated (though it is still bounded by $2\sqrt{2}$). The origin of the violation can be traced back to the non-commuting nature of the observables and the fact that the quantum state can be entangled. 

The Bell and CHSH inequalities provide tests that allow one to {experimentally} 
discriminate between LHVTs and QM. Thus they are of fundamental importance if we want to unravel the structure of the laws of nature. The violation of the CHSH inequality was established experimentally in the spin correlations of an entangled pair of photons in the famous experiments by Clauser and collaborators, Aspect and collaborators, Zeilinger and collaborators, and many others~\cite{Freedman1972,Fry:1976:exp,aspect:1981:exp,PhysRevLett.49.91,PhysRevLett.49.1804,Weihs1998,zeilinger:1998:exp,zeilinger:2002:exp,Giustina:2015yza,PhysRevLett.115.250402,hensen:2015:loophole}, where the observables $A(\ba)$ and $B(\bb)$ 
correspond to measuring the spins of the photons in directions determined by the measurement settings $\ba$ and $\bb$.
QM and entanglement are thus extremely well established in the context of quantum optics and photon spin correlations, and rule out an interpretation in terms of {an} LHVT.

Proposals to measure entanglement and non-locality at collider experiments are not new~\cite{tornqvist:1981:suggestion,tornqvist:1986:decay}. However, very recently, it was proposed that it is possible to test for entanglement versus non-entanglement at very high energies at the LHC~{\cite{Fabbrichesi:2021npl, Afik:2021:entanglement}}, which has spurred follow-up experimental activities at the ATLAS and CMS experiments~\cite{ATLAS:2024:entanglement,CMS:2024:entanglement,CMS:2024zkc} and also a very large number of phenomenological studies~\cite{Fabbrichesi:2023:Constraining,Afik:2021:entanglement,aguilar:2023:postdecay,cheng:2025:optimizing,Severi:2021cnj,Aoude:2022imd,Afik:2022kwm,Aguilar-Saavedra:2022uye,Afik:2022dgh,Severi:2022qjy,Dong:2023xiw,Cheng:2023qmz,Aguilar-Saavedra:2024fig,Maltoni:2024tul,Aguilar-Saavedra:2024hwd,Aguilar-Saavedra:2024vpd,Maltoni:2024csn,White:2024nuc,Han:2023fci,Dong:2024xsb,Fabbrichesi:2024:Quantum,Altakach:2023:Quantum,ehataeht:2024:probing,Breuning:2024wxg,LoChiatto:2024dmx,Privitera:1992:Decay,han:2025:entanglement,ma:2024:testing,zhang:2025:entanglement,Kats:2024jmy,Afik:2025grr,Barr:2024djo},
which fall into different categories: studies of angular correlations of top quark decays~\cite{Fabbrichesi:2023:Constraining,Afik:2021:entanglement,aguilar:2023:postdecay,cheng:2025:optimizing,Severi:2021cnj,Aoude:2022imd,Afik:2022kwm,Aguilar-Saavedra:2022uye,Afik:2022dgh,Severi:2022qjy,Dong:2023xiw,Cheng:2023qmz,Aguilar-Saavedra:2024fig,Maltoni:2024tul,Aguilar-Saavedra:2024hwd,Aguilar-Saavedra:2024vpd,Maltoni:2024csn,White:2024nuc,Dong:2024xsb,Han:2023fci}, $\tau$ lepton decays~\cite{Fabbrichesi:2023:Constraining,Altakach:2023:Quantum,ehataeht:2024:probing,Fabbrichesi:2024:Quantum,Breuning:2024wxg,LoChiatto:2024dmx,Privitera:1992:Decay,han:2025:entanglement,ma:2024:testing,zhang:2025:entanglement} and $b$ quark decays~\cite{Kats:2024jmy,Afik:2025grr}. {The
decays of a Higgs boson to a pair of vector bosons, $H\to VV^*$, with the vector bosons decaying 
leptonically have also been investigated~\cite{Barr:2021zcp, 
Aguilar-Saavedra:2022wam, 
Ashby-Pickering:2022umy, 
Aguilar-Saavedra:2022mpg, Fabbrichesi:2023cev, 
Aoude:2023hxv,
Bernal:2023ruk,
Bi:2023uop,
Aguilar-Saavedra:2024whi,
Subba:2024mnl,
Ruzi:2024cbt,
Grossi:2024jae,
Sullivan:2024wzl,
Wu:2024ovc}, see also the related work ref.~\cite{Grabarczyk:2024wnk}.}
In further proposals Kaons~\cite{Benatti:1998:Bell,benatti:2000:direct,Bertlmann:2001:bell,banerjee:2016:quantum}, B mesons~\cite{Fabbrichesi:2023idl,Go:2007:measurement,banerjee:2016:quantum,Go:2004:Observation} 
and charmonium decays into strange baryons~\cite{tornqvist:1981:suggestion, 
tornqvist:1986:decay, Baranov:2008:bell,baranov:2009:Bell} have been 
considered. Furthermore, critical analyses of the specific logic of such 
interpretations have been put forward~\cite{Bertlmann:2004:violation,Li:2024luk}. We refer to ref.~\cite{Barr:2024djo} for an overview of various proposals. Very recently, it 
has been claimed that entanglement has been measured with high significance at BESIII in strange baryon decays~\cite{BESIII:2025vsr}. Most of these proposals are based on measuring angular distributions and/or angular correlations of decay products, and of relating these angular observables to spin states. Some ideas (e.g.~refs.~\cite{Bertlmann:2001:bell,Go:2004:Observation}) employ flavor states and their following self-analyzing decay as a laboratory for entanglement. While the majority of our analysis will concentrate on angular observables, we will briefly comment on  tests of locality based on flavor states in the summary.

These proposals offer novel possibilities for collider phenomenology. For example, since quantum entanglement has been traditionally studied in the context of Quantum Information Theory, this may open the door to import concepts from this field and to apply them to build new collider observables for precision measurements of the Standard Model (SM) of particle physics and/or to constrain signals of new physics. At the same time, we are confronted with new challenges, as is typically the case whenever concepts from one field are applied to another. In particular, care is needed how to perform and interpret studies of entanglement at colliders in order not to result in logical fallacies.\footnote{For example, recent literature often refers to `measuring entanglement' at colliders. This is a clear misnomer: Entanglement is a property of a quantum state, and not an observable that can be measured. {Similarly, one cannot `test Bell's inequality'. Bell 
proved a mathematical theorem which is valid for all LHVTs. One can only test for locality via Bell's inequality.}}

Given the large amount of recent activity around studying entanglement and Bell's inequalities at colliders, in the present paper we critically assess 
the possibility of using collider experiments to test QM against LHVTs, e.g. by testing {for} violations of Bell's inequalities in a collider environment. Clearly, any valid experiment that claims to test QM against LHVTs cannot at the same time implicitly assume QM from the start. This seemingly obvious fact is particularly important at particle colliders, where theoretical predictions are routinely performed in the context of Quantum Field Theory (QFT). Extreme care is thus needed that QFT and QM are not assumed at any stage in the argument if QM itself is put to the test against LHVTs at a collider. Our goal is to argue that it is currently 
{\textit{not}} possible to perform any logically coherent test of entanglement versus non-entanglement and
{locality via the Bell {or} CHSH inequalities at colliders without already assuming the validity of QM from the start. {Such an assumption} invalidates any of these proposed tests. This observation is not new. In fact, it has already been pointed out more than 30 years ago that it is not possible to test locality via a Bell's inequalities at colliders~\cite{Abel:1992kz}. In essence, ref.~\cite{Abel:1992kz} starts from the fact that at a collider we cannot measure the spins of the produced particles, but only their momenta. Due to the commuting nature of the momentum operator in QM, it is always possible to construct a LHVT that reproduces the expectation values of any given observable that is a function of momenta only~\cite{Abel:1992kz,kasday}. Hence, no violation of Bell's inequalities can be observed {in this} way. {Only by assuming QM can a non-local function be extracted from
differential cross section measurements at current colliders.}

Violations of Bell's inequalities are tightly linked to measuring correlations of products of non-commuting observables, like spin correlations. Spin correlations, however, are not directly measurable at colliders.
Recent approaches to studying entanglement at colliders have mostly been based on the idea that spin correlations can be measured indirectly at colliders, by extracting them from appropriate angular correlations between decay products. This has led to a vast literature and proposals on how to test locality via Bell's inequalities and/or to establish entanglement versus non-entanglement
at colliders, apparently evading the no-go theorem of ref.~\cite{Abel:1992kz}. The main result of this paper is a critical analysis of the idea that spin correlations can be indirectly measured at colliders and used as a means to test locality through studying entanglement 
versus non-entanglement and {potential} violations of Bell's inequalities. By studying the precise relation between spin and angular correlations at colliders, we find that these approaches 
always ultimately assume QM at some stage in the reasoning. This reveals a logical fallacy also in this indirect approach of attempting to access spin correlations through angular correlations of momenta, which invalidates the proposed tests of QM vs.\ LHVTs at colliders, and confirms the no-go theorem of ref.~\cite{Abel:1992kz} (though a certain subclass of LHVTs may still be constrained, as was also pointed out in ref.~\cite{Abel:1992kz}). We emphasize that the no-go theorem applies to
the statement about tests of locality at colliders or of entanglement versus non-entanglement. It does not apply to using observables inspired from Quantum Information Theory and entanglement (e.g., concurrence) to test various models of QFT among themselves, e.g., as a means to test the SM of particle physics against various extensions of it, while always assuming QM. 

The remainder of this paper is organized as follows: In section~\ref{sec:spin} we review existing ways of relating spin and angular correlations, both in QFT and in LHVTs. In sections~\ref{sec:discussion} and~\ref{sec:exp} we show that it is currently not possible to relate spin and angular correlations without presupposing QFT and thus QM.  This shows that it is not possible to perform any logically consistent test of locality via Bell's inequalities at colliders based on spin correlations, in agreement with the no-go theorem of ref.~\cite{Abel:1992kz}. In section~\ref{sec:constraints} we discuss how measurements of angular correlations can be turned into constraints on certain classes of LHVTs. In section~\ref{sec:conclusions} we draw our conclusions.
% !TEX root = main.tex

\section{Relating spin  and angular correlations}
\label{sec:spin}

In this section we review how to relate spin  and angular correlations at colliders. More specifically, we consider the production of a fermion pair $F_+F_-$ that subsequently decays according to 
\beq
F_+ \to f_{+,1},\ldots, f_{+,n}\,,\qquad \qquad F_- \to f_{-,1},\ldots, f_{-,n}\,.
\eeq
We assume that the widths of the fermions are small enough so that they can 
be identified as sharp resonances, which motivates to consider their 
on-shell production. Prime examples for such processes are the production 
and the decays of top or $\tau$ pairs. In the following we focus on the 
correlations associated with the first two final state particles, 
respectively, $f_{\pm,1}$.

Our goal is to relate the spin correlation matrix 
\beq\label{eq:spin_C_matrix}
C_{ij} = \langle s_+^i\,s_-^j\rangle\,,\qquad  i,j\in\{1,2,3\}\,,
\eeq
where $\bs_{\pm}=(s_{\pm}^1,s_{\pm}^2,s_{\pm}^3)$ are the directions of the spins of $F_{\pm}$ in 
their respective rest frames, to the correlation matrix 
\beq\label{eq:angle_Q_matrix}
 Q_{ij} = \langle q_+^i\,q_{-}^j\rangle\,,
\eeq
where $\bq_{\pm}$ is the direction of the momentum of $f_{\pm,1}$ in the rest frame of $F_{\pm}$. Note that we assume $|\bs_{\pm}|=|\bq_{\pm}|=1$. The combination of spin correlations that enters the CHSH inequality can be cast in a form involving the spin correlation matrix $\bC$. More precisely, eq.~\eqref{eq:CHSH} is equivalent to
\beq\label{eq:CHSH_C}
\big|(\ba-\ba')^T\bC\bb + (\ba+\ba')^T\bC\bb'\big| \le 2\,.
\eeq

\subsection{Relating spin  and angular correlations in QFT}
The relationship between spin  and angular correlations in QFT is well known, and has been extensively studied in particular in the context of top pair production~\cite{Kane:1991bg,Bernreuther:1993hq,Bernreuther:2001rq,Bernreuther:2004jv,Mahlon:1995zn,Mahlon:1996pn,Mahlon:1997uc,Parke:1996pr,Uwer:2004vp,Mahlon:2010gw,Baumgart:2011wk,Baumgart:2012ay,Baumgart:2013yra,Bernreuther:2015yna}. We will therefore be brief with the derivations and only focus on the main aspects. 
 
In QFT the differential probability for the production of the fermion pair 
$F_+{F}_-$ and its subsequent decay is proportional to the matrix element 
squared of the process. In the narrow width approximation, this matrix 
element squared factorizes into the spin-correlated 
matrix element for the production of an on-shell $F_+{F}_-$ pair, connected 
by the spin density matrix, $\Gamma_{\beta\bar\beta}^{F_\pm}$,
\beq\label{eq:M_spin}
|\cM(F_+{F}_-+X\to \{f_{+,i}\},\{{f}_{-,i}\}+X)|^2 \,\,\,\sim\,\,\, \cM(F_+{F}_-+X)_{\beta\bar{\beta}}\,\Gamma^{F_+}_{\beta\beta'}\,\Gamma^{{F}_-}_{{\bar\beta}\bar{\beta}'}\,\cM(F_+{F}_-+X)_{\beta'\bar{\beta}'}^*\,,
\eeq
where $X$ denotes additional radiation (which may need to be taken into account at higher orders in perturbation theory), and $\beta^{(\prime)}$ and $\bar{\beta}^{(\prime)}$ are the spin eigenvalues of the $F_+$ and $F_-$ along some quantization axis. 
We assume that we are inclusive over all final-state momenta except for the directions $\bq_{+}$ and $\bq_{-}$ of the decay products $f_{+,1}$ and ${f}_{-,1}$. The spin density matrices are hermitian $2\times2$ matrices, and may thus be expanded into the basis $(\mathds{1}, \vec{\sigma})$ of Pauli matrices. By rotational invariance, the most general form is
\beq\label{eq:Gamma^F} 
\Gamma^{F_+} \propto \mathds{1} + \alpha_{1}\,\bq_+\!\cdot\!\vec{\sigma}\,.
\eeq
The constant of proportionality is related to the total decay rate of $F_+$, and its precise form is irrelevant here. The constant $ \alpha_{1}$ is called the \emph{spin-analyzing power} of $F_+$. Inserting eq.~\eqref{eq:Gamma^F} into eq.~\eqref{eq:M_spin}, one arrives at the following expression for the {normalized differential cross section as 
a function of} the directions $\bq_{\pm}$~\cite{Baumgart:2013yra},
\beq\label{eq:top_master_formula}
\frac{1}{\sigma}\frac{\rd\sigma}{\rd\Omega_+\rd\Omega_-} = \frac{1+\alpha_1\,\bB_+\!\cdot\!\bq_++\bar{\alpha}_{1}\,\bB_-\!\cdot\!\bq_- +\alpha_{{1}}\bar{\alpha}_{1}\, \bq_+^T\bC\bq_-}{16\pi^2}\,,
\eeq
where $\rd\Omega_{\pm}$ are the infinitesimal solid angles in the 
respective $F_{\pm}$ rest frames,} attached to the directions $\bq_{\pm}$, 
and $\bar{\alpha}_1$ is the spin-analyzing power of ${F}_-$. The vectors 
$\bB_{\pm}$ are related to the net polarizations of $F_+$ and ${F}_-$, and 
$\bC$ is the spin correlation matrix from eq.~\eqref{eq:spin_C_matrix}. From eq.~\eqref{eq:top_master_formula} we can 
easily read off the relation between the spin  and angular correlations (see 
appendix~\ref{app:integrals}),
\beq\label{eq:QFT_angular_spin}
Q_{ij}=\langle q_+^i\,q_-^j\rangle=\int_{S^2\times S^2}\rd \Omega_+\,\rd\Omega_-\,\frac{1}{\sigma}\frac{\rd\sigma}{\rd\Omega_+\rd\Omega_-}\,{q}_{+}^i\,{q}_{-}^j = \frac{\alpha_1\bar{\alpha}_{1}}{9}\,C_{ij}\,.
\eeq
Equation~\eqref{eq:QFT_angular_spin} is the desired relation between the spin  and angular correlations in QFT. This relation is well known, and is the basis for many spin correlation studies at colliders, e.g., in the context of top or $\tau$ physics. We stress that the derivation of this relation deeply relies on the use of QFT, in particular on the squared matrix element in the narrow-width approximation, which is a quantity intrinsic to QFT. Before we discuss what this relation becomes in a LHVT, let us make {a} comment about the spin-analyzing powers  $\alpha_1$ and $\bar{\alpha}_1$. It follows from CPT-invariance (which is a very general property of a local relativistic QFT) that the spin-analyzing powers of $F_+$ and ${F}_-$ are related by
\beq\label{eq:alpha_CPT}
 \bar{\alpha}_1 = - \alpha_{{1}}\,.
\eeq
Second, the spin-analyzing power is closely related to the {differential} decay rate of $F_+$ in the direction of $f_{+,1}$,
\beq
\frac{1}{\Gamma}\frac{\rd\Gamma\phantom{_+}}{\rd\Omega_+} = \frac{1+\alpha_1\,P_{+}\cos\theta}{4\pi}\,,
\eeq
where $\cos\theta=\bq_+\!\cdot\!\bd$ is the angle between $\bq_+$ and some 
reference direction $\bd$, with $|\bd|=1$, and $P_{+}$ is the polarization 
of $F_+$ in the direction $\bd$. We have
\beq
P_{+} \equiv P(\bs_+\!\cdot\!\bd>0) - P(\bs_+\!\cdot\!\bd<0)\propto \bB_+\!\cdot\!\bd\,,
\eeq
where $P(\bs_+\!\cdot\!\bd>0)$ is the probability that the spin $\bs_+$ of 
$F_+$ in the latter's rest frame points in the direction $\bd$. For a 
sample perfectly polarized in the direction $\bd$, i.e. $P_{+}=1$, we infer 
the probability for the particle $f_{+,1}$ to be emitted in the direction 
$\bq_+$, if the spin of $F_+$ points in the direction $\bs_+$ as
\beq\label{eq:P(u|s)_QFT}
 P(\bq_+|\bs_+) = \frac{1+\alpha_1\bq_+\!\cdot\!\bs_+}{4\pi}\,.
\eeq
 
\subsection{Relating spin  and angular correlations in LHVTs}
\label{sec:spin_ang_LHVTs}
We now discuss how the analysis of the previous section changes in the context of a LHVT. Since our goal is to test QM against LHVTs, we pay particular attention not to assume any concepts from QFT. Our analysis only relies on the following assumptions:
\begin{enumerate}
\item We assume a LHVT, with hidden variables $\lambda\in\Lambda$.
\item Special relativity holds, in particular Poincar\'e-invariance.
\item The particle species $F_+$ and its anti-particle ${F}_-$ have spin $\frac{1}{2}$. 
In particular, {the spin's} projection along any axis can only take the values $\pm\frac{1}{2}$.
\item  $F_+$ and ${F}_-$ decay independently, and the decays of one fermion do not depend on the measurement process of the other.
While this property is harder to check and may depend on the measurement setup, it is essential in order to perform local Bell-type 
experiments.
\end{enumerate}
Let us make a comment on CPT-invariance.
If we assume CPT-invariance, then the anti-particle ${F}_-$ has the same decay channels, lifetime and branching ratios as the particle $F_+$. CPT-invariance is automatically satisfied in the context of local, relativistic QFTs. Since our goal is to investigate if we can test QM, and thus QFT, against LHVTs, we cannot logically take CPT-invariance for granted anymore. We therefore do not assume CPT-invariance in our discussion, but we comment on it at the end. 

We start from the representation of the matrix of angular correlations, cf.
eq.~\eqref{eq:angle_Q_matrix},
\beq\label{eq:Q_LHVT}
Q_{ij} = \langle q^i_+\,q^j_-\rangle = \int_{S^2\times S^2}\rd \Omega_+\rd\Omega_-\,q^i_+\,q^j_-\,P(\bq_+,\bq_-)\,.
\eeq
{Here} $P(\bq_+,\bq_-)$ is the joint PDF to produce the final-{state} decay products, {where} 
the momentum of $f_{\pm,1}$ points in the direction $\bq_{\pm}$ in the rest frame of $F_{\pm}$. 
Let us make a comment: at first glance, the integral in eq.~\eqref{eq:Q_LHVT} does not match 
the structure of an expectation value in a LHVT as given in eqs.~\eqref{eq:A_LHVT} 
and~\eqref{eq:AB_LHVT}. In particular, eq.~\eqref{eq:Q_LHVT} does not involve any integration 
over the hidden variables $\blambda$. In appendix~\ref{app:LHVTs} we show that both 
representations are valid and follow from the axioms of a LHVT, cf.~ref.~\cite{plato}.

We now use Bayes' law to write
\beq
P(\bq_+,\bq_-) = \int_{S^2\times S^2}\rd\Omega_+^s\rd\Omega_-^s\,P(\bq_+,\bq_-|\bs_+,\bs_-)\,P(\bs_+,\bs_-)\,,
\eeq
where $\bs_{\pm}$ is the direction of the spin of $F_{\pm}$ in the 
particle's rest frame ($|\bs_{\pm}|=1$), and $\rd\Omega_{\pm}^s$ is the 
infinitesimal spherical {solid} angle associated {with} the direction 
$\bs_{\pm}$. Here $P(\bs_+,\bs_-)$ is the joint PDF to produce the final 
state $F_+F_-$ with the spins in their respective rest frames pointing in 
the directions $\bs_+$ and $\bs_-$, and $P(\bq_+,\bq_-|\bs_+,\bs_-)$ is the 
conditional probability that $F_+$ and ${F}_-$ with these polarizations 
decay into the final states with the given directions $\bq_{\pm}$ for 
$f_{+,1}$ and ${f}_{-,1}$. We may now use the assumption that the decays of 
the two fermions are independent, and that they do not depend on the 
properties of the respective other fermion. This implies
\beq\label{eq:P_fact}
P(\bq_+,\bq_-|\bs_+,\bs_-) 
= P(\bq_+|\bs_+)P(\bq_-|\bs_-)\,.
\eeq
Hence,
\beq\label{eq:Q_LHVT_2}
\langle q^i_+\,q^j_-\rangle = \int_{S^2\times S^2}\rd \Omega_+^s\rd\Omega_-^sP(\bs_+,\bs_-)\,S^i_+(\bs_+)\,S^j_-(\bs_-)\,,
\eeq
where we defined
\beq\label{eq:S_integral}
\bS_{\pm}(\bs_{\pm}) {\;\equiv} \int_{S^2}\rd\Omega_{\pm}\,\bq_{\pm}\,P(\bq_{\pm}|\bs_{\pm})\,.
\eeq

It seems that we need an explicit expression for $P(\bq|\bs)$ to proceed. In
QFT, it is given by eq.~\eqref{eq:P(u|s)_QFT}. Here, however, we cannot 
rely on QFT, and we are limited to using the assumptions spelled out at the 
beginning of {this} section. As we now show, the explicit functional 
dependence of $P(\bq|\bs)$ is to a large extent irrelevant.

By rotational invariance, we know that $P(\bq|\bs)$ can only depend on 
\beq
\cos\theta\equiv\bq\cdot\bs\,.
\eeq
Rotational invariance also implies that 
\beq\label{eq:rot_inv}
P(\mathbf{R}\bq|\mathbf{R}\bs)=P(\bq|\bs)\,, 
\eeq
for every rotation $\mathbf{R}\in \mathrm{SO}(3)$. This in turn implies 
that the functions $\bS_{\pm}$ defined in eq.~\eqref{eq:S_integral} have 
the property
\beq\label{eq:S_prop}
\bS_{\pm}(\mathbf{R}\bs_{\pm}) = \mathbf{R}\bS_{\pm}(\bs_{\pm})\,.
\eeq
Indeed, changing integration variables from $\bq_{\pm}$ to $\mathbf{R} 
\bq_{\pm}$ and using the fact that the integration measure is invariant 
under rotations, we find
\beq
\bS_{\pm}(\mathbf{R}\bs_{\pm}) = 
\int_{S^2}\rd\Omega_{\pm}\,\bq_{\pm}\,P(\bq_{\pm}|\mathbf{R}\bs_{\pm})
= 
\int_{S^2}\rd\Omega_{\pm}\,\mathbf{R}\bq_{\pm}\,P(\mathbf{R}\bq_{\pm}|\mathbf{R}\bs_{\pm}) = \mathbf{R}\bS_{\pm}(\bs_{\pm})\,.
\eeq
It is easy to see that eq.~\eqref{eq:S_prop} forces the most general form 
of $\bS_{\pm}(\bs_{\pm})$ to be 
\beq
\bS_{\pm}(\bs_{\pm}) = g_{\pm}(\bs_{\pm}^2)\,\bs_{\pm}\,,
\eeq
for some scalar functions $g_{\pm}$. Moreover, since $\bs_{\pm}^2=1$, we 
see that 
\beq\label{eq:P_integral_result}
\bS_{+}(\bs_{+}) = A\,\bs_{+}\textrm{~~~and~~~} \bS_{-}(\bs_{-}) = \bar{A}\,\bs_{-}\,,
\eeq
for some real constants $A\equiv g_+(1)$ and $\bar{A}\equiv g_-(1)$.

Inserting eq.~\eqref{eq:P_integral_result} into eq.~\eqref{eq:Q_LHVT_2}, we 
find
\beq\label{eq:spin_ang_LHVT}
Q_{ij}=\langle q^i_+\,q^j_-\rangle = A\,\bar{A}\,\int_{S^2\times S^2}\!
\rd \Omega_+^s\rd\Omega_-^sP(\bs_+,\bs_-)\,s^i_+\,s^j_-=A\,\bar{A}\,\langle s^i_+\,s^j_-\rangle = A\,\bar{A}\,C_{ij}\,.
\eeq
We see that, very generally, the matrices of angular and spin correlations 
are proportional to each other. We stress that this feature is generic, and 
does not rely at all on the detailed features of the underlying LHVT. 
Indeed, the only assumptions that went into the derivation of 
eq.~\eqref{eq:P_integral_result} are rotational invariance and the 
independence of the two measurements, expressed through 
eqs.~\eqref{eq:rot_inv} and~\eqref{eq:P_fact}, respectively.

Let us conclude by connecting the constants $A$ and $\bar{A}$ to the 
spin-analysing powers $\alpha_1$ and $\bar{\alpha}_1$ known from QFT. In 
order to see the connection, let us assume that the PDF $P(\cdot|\bs)$ is a 
continuous function of $\cos\theta$ (which is also the case in QFT). Since 
the unit sphere is compact, this immediately implies that $P(\cdot|\bs)$ is 
bounded everywhere on the unit sphere, $|P(\cdot|\bs)|\le M<\infty$, for 
some positive real number $M$ (that may depend on $\bs$). Then we have
\beq
\int_{S^2}\rd\Omega\,|P(\bq|\bs)|^2 \le 4\pi M^2< \infty\,.
\eeq
Hence, $P(\bq|\bs)$ is a square-integrable function on the sphere $S^2$. 
Thus it can be expanded into the basis of spherical harmonics. Since $P(\bq|
\bs)$ only depends on $\cos\theta$, this expansion reduces to an expansion 
in Legendre polynomials. We then have
\beq\label{eq:Legendre}
P(\bq|\bs) = \sum_{\ell=0}^\infty \frac{\alpha_{\ell}}{4\pi}\,P_{\ell}(\cos\theta)\,,
\eeq
where the $\alpha_{\ell}$ are real numbers. The normalization condition 
$\int_{S^2}\rd\Omega\,P(\bq|\bs) = 1$ implies $\alpha_0 = 1$. The other 
coefficients are left unconstrained by the normalization condition. In
appendix~\ref{app:integrals} we show that the constants $A$ and $\bar{A}$
are related to the coefficient of the Legendre polynomial with $\ell=1$,
\beq\label{eq:A_to_alpha}
A = \frac{\alpha_1}{3} \textrm{~~~and~~~} \bar{A}=\frac{\bar{\alpha}_1}{3}\,,
\eeq
%
%
%
%Having identified the most general form allowed for the function $P(\bq|\bs)$ to be an expansion in Legendre polynomials, we can insert eq.~\eqref{eq:Legendre} into eq.~\eqref{eq:S_integral}. In appendix~\ref{app:integrals} we show that
%\beq\label{eq:P_integral_result}
%S_+(\bs_+) = \frac{\alpha_1}{3}\,\bs_{+}\,, \textrm{~~~~and~~~~}S_-(\bs_-) = \frac{\bar{\alpha}_1}{3}\,\bs_{-} \,,
%\eeq
where $\bar\alpha_1$ is the coefficient of $\tfrac{1}{4\pi}\bq_-\!\cdot\!
\bs_-$ in $P(\bq_-|\bs_-)$, i.e., for $\ell=1$. Note in particular that 
the result of the integral in eq.~\eqref{eq:S_integral} is entirely 
determined by the $\ell=1$ polynomial, and it is independent of the 
coefficients of the Legendre polynomials with $\ell>1$. In analogy with the 
QFT case in eq.~\eqref{eq:P(u|s)_QFT}, we also refer to the coefficients 
$\alpha_1$ and $\bar{\alpha}_1$ as the spin-analyzing powers.

Let us make some final comments. In principle, $\alpha_\ell$ and $\bar\alpha_\ell$ 
are free parameters for $\ell>0$. However, they are naturally bounded. For example, 
we have 
\beq\label{eq:alpha_bound}
|\alpha_1| = 3|\langle\cos\theta\rangle| \le 3\langle|\cos\theta|\rangle  
= 3\int_{S^2}\rd\Omega_+\,|\cos\theta|\,P(\bq_+|\bs_+) \le 3\,.
\eeq
Similar relations hold for $\ell>1$. Note that the same bound (with exactly 
the same derivation) holds in QFT.\footnote{In QFT we have the even 
stricter bound $|\alpha_1|\le 1$.} Moreover, we have not assumed CPT-
invariance, so that a priori $\alpha_1$ and $\bar{\alpha}_1$ are distinct. 
If we impose CPT-invariance, we find
\beq
\bar{\alpha}_{\ell} = (-1)^{\ell}\alpha_{\ell}\,,
\eeq
in agreement with the corresponding result in QFT, 
cf.~eq.~\eqref{eq:alpha_CPT}.  

\section{Testing locality at colliders}
\label{sec:discussion}
Comparing eqs.~\eqref{eq:QFT_angular_spin} and~\eqref{eq:spin_ang_LHVT}, we remarkably find exactly the same relation between spin  and angular correlations in QFT and LHVTs, namely we have in both cases
\beq\label{eq:final_QC}
\bQ = \frac{\alpha_1\bar{\alpha}_1}{9}\,\bC\,.
\eeq
Let us make some comments about this relation. In fact, the final form in eq.~\eqref{eq:final_QC} has already appeared in the literature before, cf.,~e.g.,~refs.~\cite{Abel:1992kz,Altakach:2023:Quantum,Chen_master}. We believe, however, that this is the first time that a derivation has been performed that consistently only uses assumptions belonging to either QFT or LHVTs. For example, previous derivations rely on the restricted form in eq.~\eqref{eq:P(u|s)_QFT}, or they assume that the joint distribution $P(\bs_+,\bs_-)$ is always a normalized PDF. The latter assumption is not valid in QM, where $P(\bs_+,\bs_-)$ is not a PDF, but rather has to be interpreted as the Sudarshan $P$-function~\cite{PhysRevLett.10.277}, which may be negative, or even singular.

Having established eq.~\eqref{eq:final_QC} as the relation between spin  
and angular correlations in both QM and LHVTs, we may ask for the 
implications. The left-hand side of eq.~\eqref{eq:final_QC} only involves 
the matrix $\bQ$ of angular correlations, which are directly measurable at 
colliders. We can then use eq.~\eqref{eq:final_QC} to extract (and thus 
indirectly measure) the spin correlation matrix $\bC$, \emph{{however}
only if the spin-analyzing powers $\alpha_1$ and $
\bar{\alpha}_1$ are known.} Any test of locality or study 
of entanglement via spin correlations at colliders will have to face this 
fact.

Let us illustrate this on the example of the CHSH inequality. We assume
that we have measured the matrix of angular correlations in a collider 
experiment. From this we can compute the quantity 
(cf.~eq.~\eqref{eq:CHSH_C})
\beq
R_Q \equiv \max_{\ba,\bb,\ba',\bb'}\big|(\ba-\ba')^T\bQ\bb + (\ba+\ba')^T\bQ\bb'\big|\,.
\eeq
Since angular correlations do not lead to a violation of the CHSH inequality (see ref.~\cite{Abel:2025skj} for concrete examples), we have
\beq\label{eq:RQ_CHSH}
R_Q\le 2\,.
\eeq
Analogously, we define (cf.~eq.~\eqref{eq:CHSH_C})
\beq
R_C \equiv \max_{\ba,\bb,\ba',\bb'}\big|(\ba-\ba')^T\bC\bb + (\ba+\ba')^T\bC\bb'\big|\,.
\eeq
Note that, unlike $R_Q$, $R_C$ cannot be directly measured at colliders, 
but, since $R_Q$ and $R_C$ are linear in $\bQ$ and $\bC$, respectively, we 
can use eq.~\eqref{eq:final_QC} to relate $R_C$ to $R_Q$,
\beq\label{eq:RC_to_RQ}
R_C =  \frac{9}{|\alpha_1\bar{\alpha}_1|}R_Q\,.
\eeq
From eq.~\eqref{eq:alpha_bound} we see that we always have 
\beq
R_Q \le R_C\,.
\eeq
Hence, even though $R_Q$ always satisfies the CHSH bound in eq.~\eqref{eq:RQ_CHSH}, $R_C$ may violate it. Whether or not it is violated of course depends on the specific process studied, but it also depends on the model-dependent values of $\alpha_1$ and $\bar\alpha_1$, which are unknown at this point.

We then have two possibilities how to obtain values for the spin-analyzing
powers:
\begin{enumerate}
\item {\bf Theory input:} The spin-analyzing powers are computable within 
the SM (or an appropriate extension of it), and we use these theoretical 
values as input. This is what has typically been done in current 
phenomenological studies and experimental measurements. In this case, 
however, we have used a QFT computation as input to our indirect 
measurement. As a consequence, our indirect measurement presupposes QM, and 
therefore it can no longer serve as a test of QM against LHVTs. Hence, in 
this approach, any test of locality via Bell's inequalities or measurement 
that establishes entanglement at colliders is logically 
invalidated. 
\item {\bf Experimental input:} One possible way out could be to obtain the 
spin-analyzing powers from an independent direct experimental 
determination, without relying on theory input that presupposes QM. 
\end{enumerate}

We will come back to the last point in section~\ref{sec:exp}. Before that, let us illustrate the first point on the study of entanglement in two cases. First, we discuss the interpretation of the measurement of angular correlations in $e^+e^-\to\gamma\Lambda\bar\Lambda\to \gamma\,p\pi^-\,\bar p\pi^+$ from refs.~\cite{BESIII:2025vsr,BESIII:2018cnd,BESIII:2022qax} in terms of a test of entanglement, and second in top-pair production at the LHC, proposed in refs.~\cite{Fabbrichesi:2021npl,Barr:2024djo,Fabbrichesi:2023:Constraining,
Han:2023fci, cheng:2025:optimizing,aguilar:2023:postdecay} 
and measured in refs.~\cite{ATLAS:2024:entanglement,CMS:2024:entanglement}.

In the first example of ref.~\cite{BESIII:2025vsr}, BESIII interprets the 
measurement of angular correlations of protons and antiprotons in the 
process $e^+e^-\to\gamma\eta_c\to\gamma\Lambda\bar\Lambda\to \gamma\,p\pi^-
\,\bar p\pi^+$ as a measurement of entanglement and as an exclusion of 
locality in LHVTs with a significance of $5.2\,\sigma$ using Bell- and 
CHSH-type inequalities. This decay channel has already been critically 
discussed in ref.~\cite{Abel:1992kz}.\footnote{Interestingly, 
ref.~\cite{Abel:1992kz} is cited in ref.~\cite{BESIII:2025vsr}, however 
without any acknowledgment of the critical discussion.} The analysis is 
based on the idea that the expectation value $E$ of the spin correlations 
between the $\Lambda$ spins along the directions $\mathbf{n}_{1,2}$, for 
the $\Lambda$'s emerging from the $\eta_c$ decay, can be measured through 
the observation of the relative angle between the protons' momenta, 
$\theta_{p\bar p}$: 
\beq
E(\mathbf{n}_1,\mathbf{n}_2)= \langle S|\boldsymbol\sigma\cdot\mathbf{n}_1\,\boldsymbol\sigma\cdot\mathbf{n}_2|S\rangle = -\cos\theta_{p\bar p}\,,
\eeq
which is derived from QM. Furthermore, the paper by the BESIII collaboration starts with the claim that the parameter 
$\alpha_{\Lambda}$ in the function relating the measured distribution $I(\theta_{p\bar p})$ of the angle $\theta_{p\bar p}$ between the proton momenta 
\beq
I(\theta_{p\bar p}) = 1 + \alpha_{\Lambda}^2 \cos\theta_{p\bar p}\,,
\eeq
has been experimentally measured as $\alpha_{\Lambda} = 0.750 \pm 0.009 \pm 0.004$ in 
refs.~\cite{BESIII:2018cnd,BESIII:2022qax}. Although experimental input appears to 
have been used exclusively to determine the critical input parameter $\alpha_{\Lambda}$ 
independently of the assumption of QM, a look  at the derivation of the angular 
distributions from which $\alpha_{\Lambda}$ is obtained reveals that the measurement 
of $\alpha_ {\Lambda}$ {is} inferred from a fit to a QFT prediction~\cite{Faldt:2017kgy}, 
and thus it is only valid within the assumptions of QM. However, this parameter is necessary to interpret the observed distribution $I$. Therefore, the interpretation 
of the $p\bar p$ angular correlations at BESIII as a test of LHVTs 
 depends in two critical points on assuming QM in the measurement of 
decay angular parameters and thus cannot exclude LHVTs. The same is true for the next section of the paper, where the corresponding dependence on $\alpha_{\Lambda}$ appears in the CHSH inequality.

In the second example, we discuss $pp\to t\bar t+X$ production, cf. 
refs.~\cite{ATLAS:2024:entanglement,CMS:2024:entanglement, CMS:2024zkc}, where the tops 
decay leptonically and the $\bq_{\pm}$ are the directions of the momenta of the charged 
leptons in the (anti-)top rest frame, respectively. 
The distribution {of the normalized differential cross section as a function} of the 
angle $\cos\varphi=\bq_+\!\!\cdot\bq_-$ can be written as
\beq\label{eq:dcosphi}
\frac{1}{\sigma}\frac{\rd\sigma}{\rd\!\cos\varphi} = \frac{1-D^{\mathrm{Q}}\cos\varphi}{2}\,,
\eeq
where\footnote{We diverge here from the Atlas notation and introduce four 
variables $D$, $D^{\mathrm{exp}}$, $D^{\mathrm{Q}}$, $D^{\mathrm{C}}$, as 
well as $D^{\mathrm{SM}}$, below. This will hopefully help to keep the 
discussion unambiguous.} Here $D^{\mathrm{Q}} =-3\langle\cos\varphi\rangle 
^{\mathrm{Q}}=-3\Tr \bQ$ is proportional to the trace of the measured 
angular correlation matrix of the lepton momenta $\bq$.

Recently the CMS and ATLAS collaborations have measured 
angular distributions related to spin correlations in top quark decays. In 
the following, we use the measurement of ATLAS as an example. They obtained 
the value $D^{\mathrm{exp}}=-0.537 \pm0.002\,\text{(stat.)} \pm 0.019 \,
\text{(syst.)}$~\cite{ATLAS:2024:entanglement}, where $D^{\mathrm{exp}}$ is 
a measurement of $D^{\mathrm{Q}}$.

Let us examine this result in the light of the previous 
discussion. In general we can define a second variable $D^{\mathrm{C}}$ in 
terms of the spin correlation $D^{\mathrm{C}}\equiv \Tr\bC/3$. Please note the different sign and scaling between $D^\mathrm{Q}$ and $D^{\mathrm{C}}$.
%Here ``C" stands for the theory employed, which in our discussion can be QM, i.e. the SM, or an LHVT. 
Using the general connection we have established between $\bQ$ and $\bC$ in 
eq.~(\ref{eq:final_QC}), we can relate $D^{\mathrm{Q}}$ and $D^{\mathrm{C}}$
\beq
-\frac{1}{3}D^{\mathrm{Q}}\stackrel{!}{=}\langle\cos\varphi\rangle^{\mathrm{Q}} = 
\Tr \bQ = 
\frac{\alpha_1\bar\alpha_1}{9}\Tr\bC =  \frac{\alpha_1\bar\alpha_1}{3}D^{\mathrm{C}}\,.
\eeq
This equation can be read from right to left leading in the last (furthest
to the left) equality in a prediction for the observable $D^{\mathrm{Q}}$ 
based on both the spin analyzers $\alpha$ and $D^{\mathrm{C}}$. Most importantly, for the case $\alpha_1=-\bar{\alpha}_1=1$, we obtain $D^{\mathrm{Q}}=D^{\mathrm{C}}$. In QM we 
have $D^{\mathrm{SM}} \equiv \Tr \bC^{\mathrm{SM}}/3$.  Analogously the SM 
thus makes a prediction for the observed angular correlation relating 
$D^{\mathrm{Q}}$ and $D^{\textrm{SM}}$
\beq
-\frac{1}{3}D^{\mathrm{Q}}\stackrel{!}{=}\langle\cos\varphi\rangle^{\textrm{SM}} = \Tr \bQ^{\textrm{SM}}=\frac{\alpha_1^{\textrm{SM}}\bar\alpha_1^{\textrm{SM}}}{9}\Tr\bC^{\textrm{SM}} =  \frac{\alpha_1^{\textrm{SM}}\bar\alpha_1^{\textrm{SM}}}{3}D^{\textrm{SM}}=-\frac{\big(\alpha_1^{\textrm{SM}}\big)^2}{3}D^{\textrm{SM}}\,.
\label{eq:chain-SM}
\eeq
In the last equality on the right we assumed CPT-invariance. The 
spin-analyzing power of top quarks is predicted by the SM to be $|\alpha_1 
^{\mathrm{SM}}|\simeq 1$ and thus we have the prediction $D^{\mathrm{Q}} 
\simeq D^{\textrm{SM}}$. However this \textit{only} holds as such within 
the SM. In the ATLAS paper \cite{ATLAS:2024:entanglement} only 
one symbol $D$ is used throughout corresponding to setting $\alpha_1= 
\alpha^{\mathrm{SM}}=1$ and thus $D=D^{\mathrm{Q}}=D^{\textrm{SM}}$. 

Furthermore, the SM predicts the 
inequality~\cite{Afik:2021:entanglement} 
\beq\label{eq:D_condition}
D^{\textrm{SM}}<-\frac{1}{3}\,,
\eeq
based on the Peres-Horodecki criterion \cite{Peres:1996dw, 
Horodecki:1997vt} for entanglement.
The condition eq.~\eqref{eq:D_condition} can be interpreted as a sign of spin entanglement 
in top-pair production, because eq.~\eqref{eq:D_condition} is equivalent to $\langle\bs_+\!\!
\cdot\bs_-\rangle = \Tr\bC^{\textrm{SM}} =3 D^{\mathrm{SM}}< -1$. 
In any LHVT, however, the Cauchy-Schwarz 
inequality implies
\beq
|\langle\bs_+\!\!\cdot\bs_-\rangle| \le \int_{S^2\times S^2}\rd\Omega^s_+\rd\Omega^s_-\,|\bs_+\!\!\cdot\bs_-|\,P(\bs_+,\bs_-) \le \int_{S^2\times S^2}\rd\Omega^s_+\rd\Omega^s_-\,\,P(\bs_+,\bs_-) = 1\,.
\eeq
Note that this reasoning crucially relies on $P(\bs_+,\bs_-)$ being a 
normalized PDF, which is the case {for} a LHVT, but not 
necessarily in QM. Thus in an LHVT we have $|D^{\mathrm{C}}|=|\Tr\bC/3|\leq 1/3$, and therefore the $D^{\mathrm{C}}$ resulting from an LHVT always violates eq.~\eqref{eq:D_condition}. 

If the experiment had measured a correlation of spins, the Peres-Horodecki criterion would allow to make a statement about entanglement. However, the experiment has measured the angular correlation of momenta $\bq$ corresponding to
\beq
\langle\cos\varphi\rangle^{\mathrm{Q}} = \Tr \bQ\,.
\eeq
We see that ATLAS has not measured $D^{\textrm{SM}}$ directly, but only indirectly 
via eq.~(\ref{eq:chain-SM}). Thus care has to be taken to interpret this measurement. There are 
two possibilities:
\begin{enumerate}
\item
The measured value  agrees with the SM prediction,  $\langle\cos\varphi\rangle^ {\textrm{exp}}\simeq\langle\cos\varphi\rangle^{\textrm{SM}}$ (within uncertainties). 
Then it is no surprise that we have the inequality
\beq
D^{\textrm{exp}} \equiv -3\langle\cos\varphi\rangle^{\textrm{exp}} = -3\langle\cos\varphi\rangle^{\textrm{Q}}\simeq-3\langle\cos\varphi\rangle^{\textrm{SM}}= \big(\alpha_1^{\textrm{SM} }\big)^2D^{\textrm{SM}} < -\frac{1}{3}\,.
\eeq
This is not a test of entanglement versus non-entanglement but just a consistency test of the SM.
It rests on the assumption that the SM relation $|\alpha_1^{\textrm{SM}}|\simeq1$ must hold. We 
would like to stress that the fact that measured angular distributions between different momenta in top decays agree with the SM prediction cannot be taken as a proof that correlations between spins and momenta agree with the SM.
\item If instead a significant deviation between $\langle\cos\varphi\rangle^{\textrm{exp}}$ and $\langle\cos\varphi\rangle^{\textrm{SM}}$ was observed, then we may have discovered new physics in the production and/or decay of top quarks. In that case, it is logically flawed to assume $|\alpha_1^{\textrm{SM}}|\simeq1$ as a property of top decays.
\end{enumerate}
Hence, we see that, either way, we do not arrive at the conclusion that we can {unequivocally} establish entanglement in top production without presupposing QM, and so it is not logically possible to conclude that ATLAS has observed entanglement without presupposing the SM. The same applies for the methodically similar results of ref.~\cite{CMS:2024:entanglement} from the CMS collaboration.  CMS obtained $D = -0.480^{+0.016}_{-0.017}\,\text{(stat.)}^{+0.020}_{-0.023}\,\text{(syst.)}$ at the parton level.

\section{Direct measurements of the spin-analyzing power}
\label{sec:exp}

In the previous section we have argued that we cannot determine spin 
correlations from angular correlations of momenta at colliders, unless 
we use the spin-analyzing powers $\alpha_1$ and $\bar{\alpha}_1$ as inputs. If 
the spin-analyzing powers are used as theory inputs derived under the 
assumption of QM, then the spin correlations cannot provide any valid test of 
QM against LHVTs. This leaves us with the logical possibility that we may 
directly measure the spin-analyzing powers in an independent fashion, without 
assuming QM. In this section we discuss this possibility. 

Since the spin-analyzing power of $F_+$ is closely connected to the angular 
distribution of its decay products, we can try to measure $\alpha_1$ from 
the differential decay rate of $F_+$. We have in mind a situation where 
$F_{\pm}$ are charged $\tau^{\pm}$ leptons, but we keep the discussion 
general. The polarization of our sample of $F_+$'s in the beam direction 
$\bzh$ is 
\beq
P_{+} = P(\bs_+\!\cdot\!\bzh>0) - P(\bs_+\!\cdot\!\bzh<0) = \frac{N(\bs_+\!\cdot\!\bzh>0) - N(\bs_+\!\cdot\!\bzh<0)}{N(\bs_+\!\cdot\!\bzh>0) + N(\bs_+\!\cdot\!\bzh<0)}\,,
\eeq
or equivalently
\beq\label{eq:prior}
P(\bs_+\!\cdot\!\bzh>0) = \frac{1+P_{+}}{2}\,,\textrm{~~~and~~~}P(\bs_+\!\cdot\!\bzh<0) = \frac{1-P_{+}}{2}\,.
\eeq
If $\cos\theta\equiv\bq_{+}\!\cdot\!\bzh$ is the cosine of the angle between $\bq_+$ and the beam direction $\bzh$, we have
\beq\bsp\label{eq:exp_ansatz}
\frac{1}{\Gamma}\frac{\rd\Gamma}{\rd\!\cos\theta} &\,= 2\pi\,P(\bq_+|\bzh)\,P(\bs_+\!\cdot\!\bzh>0)+2\pi\,P(\bq_+|-\bzh)\,P(\bs_+\!\cdot\!\bzh<0)\\
&\,=\sum_{m=0}^{\infty} \frac{\alpha_{2m}}{2}\,P_{2m}(\cos\theta) + \sum_{m=0}^{\infty} \frac{P_{+}\alpha_{2m+1}}{2}\,P_{2m+1}(\cos\theta)\,.
\esp\eeq
We used the general form of $P(\bq_+|\bzh)$ derived in 
section~\ref{sec:spin_ang_LHVTs}, since we do not want to make any 
assumption of QM vs. LHVT at this stage. The angular distribution on the 
left-hand side is measurable at colliders. A measurement strategy could be 
to fit the distribution to a (truncated) ansatz in terms of Legendre 
polynomials, as in the last line of eq.~\eqref{eq:exp_ansatz}. However, we 
can only isolate the even coefficients $\alpha_{2m}$ from the fit, while for 
the odd Legendre coefficients we can only extract the product  $P_{+} 
\alpha_{2m+1}$ with the polarization. In particular, we are only able to 
extract the product $P_{+}\alpha_{1}$ from the fit. 

One may wonder if this is a limitation of our measurement strategy, and if we 
could find another observable that allows us to extract the value of 
$\alpha_1$ without knowledge of the polarization $P_{+}$. We believe that 
this is not possible, as we now explain. At current collider experiments, we 
only have access to the momenta of the particles; there is 
no direct spin measurement. The spin-analyzing power is connected to the conditional probability $P(\bq_+|\bzh)$, which requires the knowledge of a spin quantization axis $\bzh$. Connecting a measurement of only momenta to the spin-analyzing power requires Bayes' law as in the first line of eq.~\eqref{eq:exp_ansatz}. The prior for the spin direction in eq.~\eqref{eq:prior} then necessarily involves the polarization. Hence, we will not be able to extract $\alpha_1$ independently from $P_+$. 

This leaves us with the possibility to determine the polarization by an independent experimental measurement.
Note that the measurements of the $\tau$ lepton polarization $P_\tau$ at the LHC in $pp\to Z\to\tau\tau$ events~\cite{ATLAS:2017xuc,CMS:2023mgq} do not qualify as an independent measurement of $P_\tau$, because they use the QFT prediction for the decay angular distribution  of a variety of $\tau$ decay final states as an input to the interpretation of the measurement. 
Then, the only way to know the $\tau$ polarization $P_\tau$ independently from the  $\tau$ decay is to determine it in the production of the $\tau$. At first glance this seems indeed possible when polarized beams are used in an $e^+e^-$ collider. Assuming an idealized beam polarization of $P_{e^-}=-1$ and $P_{e^+}=+1$, the polarization of the $\tau$ pair would be given by the scattering angle $\theta_\tau$ of the $\tau$ with respect to the beam axis as in $P_+=P_\tau=\cos\theta_\tau$. This result is independent of QM itself and can be derived only from assuming angular momentum conservation and projecting the spin of the initial state onto the final state. Then, in bins of $\cos\theta_\tau$, $P_+$ would be known. Even when using realistic polarizations for, e.g., a future $e^+e^-$ linear collider of $P_{e^-}=-0.8$ and $P_{e^+}=+0.3$, the remaining polarization of the $\tau$'s might result in sufficient statistical precision on $P_+$ to disentangle $P_+$ and $\alpha_1$ in the decay angular distributions.

However, the above proposal does not stand up to scrutiny, because such a 
measurement of the $\tau$ polarization also presupposes QM, albeit 
indirectly. It is derived from the beam polarization, and indeed, all known 
ways of measuring the beam polarization (see ref.~\cite{List:2020wns} for an 
overview and ref.~\cite{Marchesini:2011aka} for an in-depth description of the measurement from  $W^+W^-$ production) resort to testing quantum mechanical predictions: The polarimeter 
measurement is based on observing the angular distribution of Compton scattering of a polarized laser beam on the polarized beam. The polarization 
is extracted from fitting the QFT prediction of polarized Compton scattering to the data. The same method is used for the in-situ measurement in $e^+e^-\to W^+W^-\to \ell^+\nu\,\ell^-\bar\nu$, where the cross-section and angular measurements of the leptons in the final state are fitted to the predictions from QFT, and from which the polarization is then extracted together with other information about the interaction. 
We thus see that, at least in current collider experiments, it is not possible to measure the polarization $P_{+}$ independently from the combination $P_+\alpha_1$, and thus we cannot extract $\alpha_1$ directly from experiment without presupposing QM.
% !TEX root = main.tex

\section{Constraining subsets of LHVTs}
\label{sec:constraints}

In the previous sections we have argued that the only way to measure spin 
correlations in collider experiments is by extracting them from angular correlations. This procedure involves the spin-analyzing powers $\alpha_1$ and $\bar{\alpha}_1$ (and $\bar\alpha_1 = -\alpha_1$ if we assume CPT-invariance) via eq.~\eqref{eq:final_QC}, and we have argued that there is currently no way to extract the spin-analyzing powers from experiments without presupposing QM. Therefore, any collider test of locality and violation of Bell's inequalities based on spin correlations  cannot succeed in unequivocally establishing QM and ruling out LHVTs. This is in agreement with the no-go theorem of ref.~\cite{Abel:1992kz}.

In ref.~\cite{Abel:1992kz} it was also argued that it is possible to rule out certain subsets of LHVTs (see also ref.~\cite{Li:2024luk}). In the following we elaborate on this fact, using the setup of the CHSH inequality from section~\ref{sec:discussion}, from where we also borrow all the notations and conventions. For simplicity, we assume that CPT-invariance holds, though the extension of the discussion to include CPT violation is immediate. Since we cannot measure $\alpha_1=-\bar\alpha_1$ directly, we consider it a parameter of the model (which could be either a QFT or a LHVT), bounded by eq.~\eqref{eq:alpha_bound}. Our goal is to determine if we exclude LHVTs with a certain value of $\alpha_1$, given a certain experimental measurement of $R_Q$. We will consider two different approaches how to do this and evaluate their validity.

\paragraph{Attempting to exclude LHVTs based on $R_C$ instead of $R_Q$.}
The first approach consists in fixing $\alpha_1$ to some value, for example the value predicted by the SM, $\alpha_1=\alpha_1^{\textrm{SM}}$. We can then ask ourselves if we can learn anything from a value of $R_C$ derived from an experimental measurement of $R_Q$ under the assumption of the fixed value of $\alpha_1$ that we could not already have known from evaluating $R_Q$ alone. Such a test only makes sense if we assume that the LHVT agrees with the experimental measurement of $R_Q$ -- otherwise there would be no reason to base the test on $R_C$, because if a specific LHVT disagrees with a specific measured value of $R_Q$ there is already enough reason to discard that LHVT. Since according to eq.~\eqref{eq:RQ_CHSH} we always have $R_Q\leq2$, though, there can always be a set of LHVTs which satisfy the measured $R_Q$. Note that this is the approach taken in the literature in many phenomenological and experimental studies of testing locality at colliders. If the value of $R_C$ derived from the measurement does violate the CHSH inequality, it is tantalizing to speculate that we have excluded LHVTs with $\alpha_1\simeq\alpha_1^{\textrm{SM}}$. However, it turns out that from such a test no LHVT can be excluded that is not already excluded by the assumptions of the interpretation of the measurement: For it to be explained by an LHVT, said LHVT would need to satisfy $R_C\leq2$. Together with $\alpha_1=\alpha_{1}^{SM}$ and $R_Q\leq2$, this implies that we already know that no LHVT exists that both predicts the measured $R_Q$ and has $\alpha_1=\alpha_{1}^{SM}$, because that theory would predict a $R_C>2$, which no LHVT can. Thus, for every LHVT that is in agreement with the measured $R_Q$ that leads to $R_C>2$ under the assumption $\alpha_1=\alpha_1^{\textrm{SM}}$, the assumption on $\alpha_1$ cannot be justified, and thus nothing can be learned from testing against $R_C$ instead of testing against $R_Q$.

\paragraph{Deriving bounds on $\alpha_1$.} The other approach consists in using the experimental measurement to derive a bound on $\alpha_1$ so that the CHSH inequality is satisfied. It is easy to check that the condition $R_C\le 2$ reduces to
\beq\label{eq:Relation_RQ_alpha}
 3\sqrt{\frac{R_Q}{2}}\le |\alpha_1|\le3\,,
\eeq
where the upper bound comes from eq.~\eqref{eq:alpha_bound}. A measurement of $R_Q$ then allows us to exclude LHVTs which violate that bound, i.e., LHVTs with $|\alpha_1|  <  3\sqrt{\frac{R_Q}{2}}$. We stress, however, that LHVTs that satisfy the bound cannot be ruled out by collider experiments.
\medskip

We can use eq.~\eqref{eq:Relation_RQ_alpha} and check which range of $\alpha_1$ in an LHVT would make it impossible to exclude that LHVT if we measure the value of $R_Q$ predicted by the SM. The exact value of that bound depends on the specific CHSH test and the specific set of observables. As an example, we consider the analysis of $e^+e^-\to Z\to\tau\tau$ processes at $\sqrt{s}=91.2$\,GeV and $e^+e^-\to ZH\to ff\,\tau\tau$ at $\sqrt{s}=240$\,GeV studied in detail in refs.~\cite{Abel:1992kz,Breuning:2024wxg,Fabbrichesi:2024:Quantum}. For these analyses, a measurement of an observable equivalent to $R_Q$ around the SM value cannot anymore lead to a violation of the CHSH inequality for $R_C$ if $\alpha_1$ exceeds a value of $\sqrt[4]{1.238} \approx 1.055$ or $\sqrt[4]{2} \approx 1.19$, respectively, well within the theoretical bound of eq.~\eqref{eq:alpha_bound}. From this example we can observe that already a mild deviation of the spin-analyzing power of a LHVT compared to the SM can shield said LHVT from exclusion. 
% !TEX root = main.tex

\section{Discussion and conclusion}
\label{sec:conclusions}

In this paper, we have critically analyzed the possibility of testing 
locality using Bell- or CHSH-type inequalities or establishing 
entanglement versus non-entanglement in current collider experiments 
without already presupposing the SM at some point in the reasoning. If such 
an experiment were feasible, it would allow us to test QM against LHVTs in 
a high-energy, relativistic environment, which is very different from the 
quantum optics experiments in which violations of Bell's inequalities have 
been established. Moreover, quantum optic experiments are performed on 
photons, while collider-based experiments would allow us for the first time 
to test QM against LHVTs in fermionic systems, as in the 
original paper by Bell \cite{Bell:1964kc}. Most importantly, it would 
allow for a revolutionary test: Excluding LHVTs without resorting to 
employing non-commutating sets of measurements, and consequently without 
the need of a macroscopic manipulation of the measurement during the 
lifetime of the particles under study. This has spurred a great deal of 
activity in the collider physics community. Since the question of testing 
QM against LHVTs is such a fundamental one, it is important to critically 
assess the proposed experiments, to make sure that they do not present any 
fallacies or loopholes that would invalidate the conclusions.

In fact, it is well known that many, but not all (for (nearly) loophole-
free experiments see, e.g., refs.~\cite{Giustina:2015yza, 
PhysRevLett.115.250402, hensen:2015:loophole}), measurements of quantum 
non-locality suffer from a series of loopholes, caused by limitations in 
the experimental realization of idealized assumptions in the CHSH and Bell 
inequality derivations. These loopholes have mostly been put forward in the 
context of quantum optics based tests of locality, and they also apply to 
collider environments. ({S}ee, e.g., ref.~\cite{Barr:2024djo} and 
references therein for an overview of loopholes at collider experiments.) 
These loopholes are present and need to be addressed. As an example 
relevant for collider experiments, we can consider the detection 
loophole~\cite{Pearle:1970zt}. Here, it is assumed that the failure to 
detect a particle in its final state is not only a function of random 
inefficiencies of the measurement apparatus, but it is at least partly 
governed by an underlying feature of the LHVT. Particle physics experiments 
presently do not meet the requirement of detecting a 
sufficient fraction of the final states considered to exclude
this loophole. Therefore, the question arises of whether the problems 
outlined in previous sections are just another loophole that could be 
closed by improving the experimental apparatus: for example, the detection 
loophole may be closed by a gradual increase of efficiency and background 
rejection through building more and more ideal detectors. The purpose of 
this paper, however, is to address a more fundamental issue: Is it 
even possible at all to test locality or entanglement versus non-
entanglement at a collider?

It has been pointed out a long time ago~\cite{Abel:1992kz} that there are 
fundamental obstacles in testing locality in collider experiments (see also
refs.~\cite{Bertlmann:2004:violation, Li:2024luk}). Given the recent vast 
phenomenological and experimental activity, the goal of our paper is to 
perform an explicit detailed study of the no-go theorem of 
ref.~\cite{Abel:1992kz}, in order to highlight its features and to explain 
in detail the reasons why it is intrinsically not possible to test locality 
in current collider experiments. In essence, the issue boils down to our 
inability to perform direct measurements of spin correlations at 
colliders, as already pointed out in ref.~\cite{Abel:1992kz}. Instead, they 
can only be inferred from angular correlations. We have meticulously 
studied the relationship between spin and angular correlations in both QM 
and LHVT, and remarkably, the relation between the two is identical in both 
classes of 
theories (though the derivations are quite distinct). In both cases the 
relation involves the spin-analyzing powers of the decaying particles, which need to be provided independently. They do not follow from a direct measurement of the angular correlations. Hence, no test of locality or entanglement can be logically consistent if the SM is presupposed in the values used for the spin-analyzing power, which is typically the case when these values are taken from theory (e.g., by fixing them to their SM values). 

We have also studied the possibility of directly measuring the spin-analyzing powers independently in the experiment, which would be a way to obtain these values without presuming the SM. However, also in this direction one hits a roadblock. We have shown that it is possible to measure only the product of the spin-analyzing power and the polarization of the decaying particles. This is rooted (yet again) in the fact that collider experiments only measure momenta, and not spins, and the conversion from momenta to spins proceeds via Bayes' law, which introduces the product of the two quantities. An independent measurement of the polarization from the polarization of the initial state (e.g., at an $e^+e^-$ collider) is currently also not feasible without using information from theory, thus again presupposing the SM. Hence, we conclude that with current experimental setups at colliders, where we can only measure momenta and not spins, it is not possible to perform a logically coherent and independent test of locality, reaffirming the no-go theorem of ref.~\cite{Abel:1992kz}. At the same time, our analysis extends the results of ref.~\cite{Abel:1992kz} and shows that it is also not possible to establish the presence of entanglement at current collider experiments without presupposing the SM. 

We note another interesting difference between collider experiments and polarimeter-based 
experiments, such as those in refs.~\cite{Giustina:2015yza,PhysRevLett.115.250402,hensen:2015:loophole}. 
The latter critically offer the option to experimentally establish the non-commuting nature 
of spin-related measurements, because the same particle, say a photon, can be measured 
consecutively in two polarimeters at different orientations $x$ and $y$. Such an 
experimental confirmation of non-commutation only occurs in measurements where a voluntary 
macroscopic manipulation of the measurement device is possible (and required) during the 
lifetime of the particles studied in the measurement. No such possibility exists in current 
collider-based experiments. No set of non-commuting measurements can be performed, and 
consequently no repetition of a polarization measurement on the same particle is possible 
to prove the non-commuting nature of a set of measurements. We would like to add the remark 
that if it had been possible to construct the equivalent of a set of non-commutating 
measurements from a combination of commutating measurements, it would potentially have 
represented one of the greatest breakthroughs in our understanding of the measurement 
process in QM since its inception, with implications far beyond merely confirming 
non-locality in collider measurements. 

The main argument underpinning our inability to test locality at current 
collider experiments goes in essence back to Kasday~\cite{kasday}.  In 
ref.~\cite{Abel:1992kz} it was shown how to extend Kasday's work on 
correlated photons to the collider context and thus how  to construct an 
LHVT that reproduces distributions at collider experiments where only the 
momenta of the final state particles are measured. Said LHVT can obviously 
not be excluded based on the measurement outcomes alone, and so this LHVT 
is consistent with these experimental outcomes. We stress that this is very 
different from the famous result by Werner~\cite{Werner:1989zz}. In his 
seminal paper, Werner has shown that a quantum theory may admit so-called 
\emph{Bell local states}, which may be entangled, but still have the 
property that the expectation value of any product of two observables $A$ 
and $B$ admits a description in terms of an LHVT. Thus QM and the LHVT 
give the same experimental predictions on the product of these two 
observables. 

The interpretation and the implications of Kasday's construction are 
different. For example, if we assume the SM of particle physics, spin 
correlations in a specific set-up may violate CHSH-type inequalities. In 
that case, these spin correlations cannot be described by an LHVT, and 
the underlying quantum state cannot be a Bell local state. The spin 
correlations, however, are not directly measurable at current colliders, 
and need to be extracted from angular correlations using the 
model-dependent spin analyzing powers. The angular correlations in turn 
admit a description in terms of an LHVT through Kasday's construction, 
even though the quantum state is not Bell local.

We have focused in our analysis on proposed Bell or CHSH tests or studies 
of entanglement based on angular observables. However, we would like to 
stress that we expect that the no-go theorem of non-locality tests at 
colliders also applies to other measurements, such as self-analyzing flavor 
states or using measurements of CP-violation. As already noticed 
in refs.~\cite{Bertlmann:2004:violation,Bramon:2004pp}, among other 
conceptual challenges, QFT is also implicitly assumed in using the 
corresponding self-analyzing decay. In fact, we expect that the proposals 
to test locality using flavor states or CP-violation suffer from similar 
problems as discussed in this paper in the context of spin-correlations, 
and we expect them to be subject to the same no-go theorem.

We reiterate that the no-go theorem only applies to tests of locality in collider experiments or to tests that aim at establishing the existence of entanglement in a system without presupposing the SM. In particular, it does not apply to approaches that aim at transferring concepts from Quantum Information Theory to build observables to discriminate between different QFTs describing models of particle physics, e.g., between the SM and various extensions of it that by construction retain the SM spin analyzers.

Finally, we mention that our inability to test locality or entanglement versus non-entanglement at colliders is reminiscent of the Münchhausen trilemma in philosophy~\cite{HansAlbert}, which can be stated as asserting that there are only three ways of proving any absolute truth:
\begin{itemize}
\item a circular argument, where a step in a proof presupposes the 
statement one wants to prove,
\item an infinite regression, where every step in a proof requires another 
proof asserting its validity,
\item dogmatism, which consists in accepting certain statements as true 
without requiring an independent proof. 
\end{itemize}
Indeed, we have argued at length how fixing the spin-analyzing power to its 
SM value presupposes QM, thereby leading to a circular argument. The 
possibility of measuring the spin-analyzing power independently led us to 
study how polarizations of decaying particles can be measured, which 
requires one to understand how polarizations of initial states are 
determined without presupposing QM in general or the SM in particular, 
thereby leading to a regression. This only leaves dogmatism, which would 
consist in simply accepting that the spin-analyzing power or the 
polarization are known by some means without questioning its origin. We 
consider each of these possibilities equally unsatisfactory, reaffirming 
the no-go theorem of ref.~\cite{Abel:1992kz} that it is not possible to 
test locality and entanglement versus non-entanglement at colliders using 
spin correlations without already presupposing the SM.

\section*{Acknowlegdements}
We thank Klaus Desch, Christian Grefe, Kazuki Sakurai, Fabio Maltoni and Xerxes Tata for important constructive input and discussions. The Cluster of Excellence EXC~3107 \enquote{Color meets Flavor} has already played an important role in its constituting phase in bringing the authors from theory and experiment together and in supporting their collaboration. 

\appendix
% !TEX root = main.tex

\section{Local Hidden Variable Theories}
\label{app:LHVTs}

In this appendix we give a short review of LHVTs, because they feature prominently in the context of Bell's inequalities, but they are typically not introduced in detail in textbooks on quantum physics. The discussion in this appendix closely follows the exposition in ref.~\cite{plato}.

LHVTs postulate the existence of an ensemble of pairs of systems, labeled 1 and 2. Each pair 
is characterized by a set of local hidden variables $\lambda\in\Lambda$, but otherwise the 
two systems are distinct. We also assume that $\Lambda$ is equipped with a PDF $P(\lambda)$.
We can perform experiments on each system. Assume that we perform an experiment $A_i$ with 
setting $\ba_i$ on the system $i\in\{1,2\}$, and assume that each experiment may deliver a 
result $s_i\in\mathcal{S}_i\subseteq\mathbb{R}$. We assume \emph{measurement independence}, 
by which we mean that the experiment settings $\ba_i$ are independent of the local hidden 
variables $\lambda$ and of the way the original system was prepared.

We postulate that there is a joint PDF -- called \emph{response probability} -- 
$p_{\ba_1,\ba_2}(s_1,s_2|\lambda)$ that the experiment $A_i$ with setting $\vec a_i$ 
returns the value $s_i$. We also assume \emph{Bell locality}, by which we mean that 
the joint PDF factorizes as
\beq\label{eq:Bell_locality}
p_{\ba_1,\ba_2}(s_1,s_2|\lambda) = p^1_{\ba_1}(s_1|\lambda)p^2_{\ba_2}(s_2|\lambda)\,.
\eeq
Note that in this setting, experiments in an LHVT itself only deliver the result $s_i$ with a given probability, even for fixed $\lambda$. We could impose in addition \emph{outer determinism} (sometimes also called \emph{realism}), by requiring that $p_{\ba_1,\ba_2}(s_1,s_2|\lambda) $ takes values in $\{0,1\}$. For our purposes, this is not required.

We can then define the expectation value of the product in the usual way
\beq
\langle A_1(\ba_1)\,A_2(\ba_2)\rangle = \int_{\Lambda}\rd^n\lambda\int_{\mathcal{S}_1}\rd s_1\int_{\mathcal{S}_2}\rd s_2\,s_1\,s_2\,p_{\ba_1,\ba_2}(s_1,s_2|\lambda)\,P(\lambda)
=\int_{\Lambda}\rd^n\lambda\,A_1(\ba_1,\lambda)\,A_2(\ba_2,\lambda)\,P(\lambda)\,,
\eeq
where in the last step we used Bell locality in eq.~\eqref{eq:Bell_locality} to define
\beq
A_i(\ba_i,\lambda) \equiv \int_{\mathcal{S}_i}\rd s_i\,s_i\,p^i_{\ba_i}(s_i|\lambda)\,.
\eeq
We recover in this way eq.~\eqref{eq:AB_LHVT} with $A_1=A,\,A_2=B$. 
Alternatively, we can also integrate over the local hidden variables to obtain
\beq
\langle A_1(\ba_1)\,A_2(\ba_2)\rangle = \int_{\mathcal{S}_1}\rd s_1\int_{\mathcal{S}_2}\rd s_2\,s_1\,s_2\,P_{\ba_1,\ba_2}(s_1,s_2)\,,
\eeq
where we defined the joint PDF 
\beq
 P_{\ba_1,\ba_2}(s_1,s_2) = \int_{\Lambda}\rd^n\lambda\,p_{\ba_1,\ba_2}(s_1,s_2|\lambda)\,P(\lambda)\,.
 \eeq
This justifies eq.~\eqref{eq:Q_LHVT}.
% !TEX root = main.tex

\section{Some useful Integrals in rewriting $P(\bq|\bs)$}
\label{app:integrals}

The following angular integrals can be used to arrive at eq.~\eqref{eq:QFT_angular_spin}:
\beq\bsp
\int_{S^2}\rd\Omega_{\pm}\,{q}^i_{\pm} &\,=0\,,\\
\int_{S^2}\rd\Omega_{\pm}\,{q}^i_{\pm}\,{q}^j_{\pm} &\,=\frac{4\pi}{3}\,\delta_{ij}\,.
\esp\eeq

We now show that if $P(\bq|\bs)$ has the form in eq.~\eqref{eq:Legendre}, 
then necessarily
\beq
\int_{S^2}\rd\Omega\,\bq\,P(\bq|\bs) = \frac{\alpha_1}{3}\bs\,.
\eeq
Note that this implies eq.~\eqref{eq:A_to_alpha}. Let $\textbf{R}$ be the
$3\times 3$ rotation matrix such that $\bs = \textbf{R}\bzh$, where 
$\bzh=(0,0,1)$ is the unit vector pointing along the $z$-axis. Then, using 
rotational invariance
\beq
\int_{S^2}\rd\Omega\,\bq\,P(\bq|\bs) = \int_{S^2}\rd\Omega\,\textbf{R}\bq\,P(\textbf{R}\bq|\textbf{R}\bzh) = \textbf{R}\int_{S^2}\rd\Omega\,\bq\,P(\bq|\bzh)\,.
\eeq
We now choose standard spherical coordinates on $S^2$, 
\beq
\bq = (\cos\varphi\,\sin\theta,\sin\varphi\,\sin\theta,\cos\theta)\,,\qquad\qquad \bq\cdot\bzh = \cos\theta\,,\qquad\qquad \rd\Omega = \rd\!\cos\theta\,\rd\varphi\,.
\eeq
If we use the orthogonality of the Legendre polynomials,
\beq
\int_{-1}^1\rd x\,P_{\ell}(x)\,P_{\ell'}(x) = \frac{2}{2\ell+1}\delta_{\ell\ell'}\,,
\eeq
it is easy to show that
\beq
\int_{S^2}\rd\Omega\,\bq\,P(\bq|\bzh) = \frac{\alpha_1}{3}\bzh\,.
\eeq
The result immediately follows.

%%%
%%% Bibliography
%%%
% ========== ========== ========== ========== ==========
\printbibliography
\end{document}